\documentstyle[11pt,moriond,epsfig]{article}

\def\be{\begin{equation}}
\def\ee{\end{equation}}
\def\bea{\begin{eqnarray}}
\def\eea{\end{eqnarray}}

\begin{document}
\vspace*{4cm}

\title{AMPLITUDE OF SUPERHORIZON COSMOLOGICAL PERTURBATIONS}

\author{J\'ER\^OME MARTIN}

\address{DARC, Observatoire de Paris, \\ 
UPR 176 CNRS, 92195 Meudon Cedex, France. \\
e-mail: martin@edelweiss.obspm.fr}

\author{DOMINIK J. SCHWARZ}
\address{Institut f\"ur Theoretische Physik, \\
Robert-Mayer-Stra\ss e 8 -- 10, Postfach 11 19 32, \\ 
D-60054 Frankfurt am Main, Germany. \\
e-mail: dschwarz@th.physik.uni-frankfurt.de}

\maketitle\abstracts{We study the influence of reheating on 
super-horizon density perturbations and gravitational waves. 
We correct wrong claims \cite{G} about the joining of perturbations 
at cosmological transitions and about the quantization of 
cosmological perturbations.}

\section{Introduction}

The aim of these proceedings, based on Ref.~2, is to clear up the 
recent controversy on the relative contribution 
of density perturbations and gravitational waves to the CMBR today.
\par
The background model is taken to be a spatially flat FLRW model. We 
restrict our considerations to density perturbations and gravitational 
waves. The most general line element reads:
\begin{equation}
\label{0}
{\rm d}s^2 = a^2(\eta )\{- (1+2\phi){\rm d}\eta ^2 + 2B_{|i}{\rm d}x^i
{\rm d}\eta + [(1-2\psi)\gamma _{ij}+2E_{|i|j}+h_{ij}^{\rm TT}
{\rm d}x^i{\rm d}x^j\} \ . 
\end{equation}
In the synchronous gauge, without loss of generality, the same line 
element can be written as:
\begin{equation}
\label{1}
{\rm d}s^2=a^2(\eta )\{-{\rm d}\eta ^2+[(1+hQ)\delta _{ij}
+\frac{h_l}{k^2}Q_{,i,j}+h_{\rm gw}Q_{ij}]{\rm d}x^i{\rm d}x^j\}.
\end{equation}
The scalar function $Q$ satisfies the Helmholtz equation and $Q_{ij}$ 
is a symmetric, transverse and traceless spherical harmonic. $k$ is 
the comoving wave number. In the following we use this form 
of the line element to make contact with previous works \cite{G}. 
It is convenient to define ${\cal H}\equiv a'/a$, which is related 
with the Hubble constant $H$ by the relation $H={\cal H}/a$. 
A prime denotes a derivative with respect to conformal time. 
The quantity $\gamma $ is defined by the expression: 
$\gamma (\eta )\equiv 1-{\cal H}'/{\cal H}^2$. For the de Sitter space-time 
$\gamma$ vanishes.
\par
Let us now consider the equations of motion for the perturbed metric. For 
density perturbations (in the case of a vanishing anisotropic 
pressure), all relevant quantities can be expressed 
in terms of the variable $\mu $ defined by: 
$\mu \equiv a(h'+{\cal H}\gamma h)/({\cal H} \sqrt{\gamma })$, except 
in the de Sitter case, which must be treated separately. For 
gravitational waves, the relevant quantity is $\mu _{\rm gw}\equiv 
ah_{\rm gw}$. Using the perturbed Einstein equations, one can show that 
both types of perturbations obey the same class of equation, 
i.e. the equation of a parametric oscillator: 
\begin{equation}
\label{mu}
\mu ''+[k^2-U(\eta )]\mu =0 \ ,
\end{equation}
with $U_{\rm dp}=(a\sqrt{\gamma })''/(a\sqrt{\gamma })$ 
and $U_{\rm gw}=a''/a$. But there is a fundamental 
difference: the presence of the factor $\sqrt{\gamma }$ in the effective 
potential of density perturbations.
\par
Equation (\ref{mu}) is valid for any model. However, it is important 
to consider cases where exact analytical solutions can be found. This  
happens for power law inflation where the scale factor 
is given by $a(\eta ) =l_0|\eta |^{1+\beta }, \beta \le -2$. We 
will consider a model with three epochs in succession: inflation, 
radiation-dominated era, and matter-dominated era. 
For this simple model, the function 
$\gamma (\eta )$ is a constant during each epoch and is given by: 
$\gamma_{\rm i}=(2+\beta )/(1+\beta ), \gamma_{\rm r}=2, 
\gamma_{\rm m}=3/2$. Therefore the constant factor $\sqrt{\gamma }$ 
drops out of $U_{\rm dp}(\eta )$, 
and the solutions for density perturbations and gravitational waves 
are given by the same Bessel functions: 
\begin{equation}
\label{10}
\mu (\eta )=(k\eta )^{1/2}[A_1J_{\beta +1/2}(k\eta )+
A_2J_{-(\beta +1/2)}(k\eta )].
\end{equation}
The aim is now to compute the amplitude of both types of perturbations 
during the matter-dominated era. In order to perform this calculation, 
two questions must be addressed:
\par
1) The initial conditions must be fixed. This amounts to choose 
the coefficients $A_1$ and $A_2$. This will be done with the help 
of quantum mechanical considerations.
\par
2) The way the solutions are matched between different epochs must be 
specified. This is no problem for gravitational waves since the 
effective potential is well-defined (although discontinuous). This is 
more tricky for density perturbations. $\gamma $ is a Heaviside function.
This means that the effective potential of the density perturbations 
is not defined in the sense of distributions at the 
different transitions. Recently, there was a controversy 
on this point \cite{G,DM}. One purpose of Ref.~2 and this paper is 
to clear up this question. This question arises because we consider 
simple models which allow analytical solutions. In reality, the 
transition is smooth and the effective potential is never ill-defined. 
  
\section{Determination of the initial conditions}

Let us start with the first question. The normalization 
of the perturbed scalar field is fixed by the uncertainty principle of 
Quantum Mechanics. In the high frequency regime, this leads to:
\begin{equation}
\label{11}
\delta \hat{\varphi }(\eta ,{\bf x})= \frac{1}{(2\pi )^{2/3}}\int 
{\rm d}{\bf k}\hat{\varphi }_1(\eta ,{\bf k})e^{i{\bf k}\cdot {\bf x}}=
\frac{\sqrt{\hbar c}}{a(\eta )}
\frac{1}{(2\pi )^{3/2}}\int \frac{{\rm d}{\bf k}}{\sqrt{2k}}
[c_{\bf k}(\eta )e^{i{\bf k}\cdot {\bf x}}+c_{{\bf k}}^{\dagger }(\eta )
e^{-i{\bf k}\cdot {\bf x}}],
\end{equation}
where the annihilation and creation operators satisfy the usual commutation 
relation. A hat indicates that we are now dealing with operators. The 
normalization of the scalar perturbations is fixed by the normalization of 
the perturbed scalar field since they are linked through Einstein's 
equations. In the high frequency limit, this link is expressed as:
\begin{equation}
\label{12}
\lim _{k \rightarrow +\infty }\hat{\mu }(\eta ,{\bf k})=-\sqrt{2\kappa }a
\hat{\varphi }_1(\eta ,{\bf k}),
\end{equation}
where $\kappa \equiv 8\pi G$. For the model under considerations, it is 
easy to find the time dependence of the creation and annihilation 
operators. As a consequence all quantities are fixed uniquely, 
if at some initial time $\eta _0$ the scalar field is placed in the 
vacuum state $c_{{\bf k}}(\eta =\eta _0)|0\rangle =0$. For gravitational 
waves the line of reasoning is similar. The result\cite{MS} is that the A's 
satisfy $|A_1^{\rm (gw)}|=|A_1^{\rm (dp)}|=|A_2^{\rm (gw)}|
=|A_2^{\rm (dp)}| = (2\pi l_{\rm Pl})/(|\cos(\pi\beta)|\sqrt{k})$. 
We will show below that $|A_1^{\rm (gw)}/A_1^{\rm (dp)}|=1$ 
turns out to be crucial to  predict the relative contribution of both 
types of perturbations.

\section{The junction conditions}

Let us now turn to the second question, i.e. the junction conditions. 
We analyze the approach taken by Grishchuk\cite{G} and compare it with the 
claims of Deruelle and Mukhanov\cite{DM}. Suppose that the spatial transition 
hypersurface $\Sigma$ is defined by its normal $n^{\mu }$. To join 
two space-time manifolds along $\Sigma $ without a surface layer
the induced spatial metric $h_{ij}\equiv g_{ij}+n_in_j$ and the extrinsic 
curvature $K_{ij}\equiv (-1/2){\cal L}_nh_{ij}$ should be continuous on 
$\Sigma$. In order to compute $K_{ij}$ the system of coordinates 
(i.e. the gauge) and the vector $n^{\mu }$ (i.e. the surface of transition) 
have to be specified. Different choices for $n^{\mu }$ lead to inequivalent 
junction conditions. 
If the surface of transition is defined by $q(\eta, x^i)=cte$, 
it was shown 
by Deruelle and Mukhanov \cite{DM} that the matching conditions read:
\begin{eqnarray}
\label{16}
& & [\psi +{\cal H}\frac{\delta q}{q_0'}]_{\pm }=0, \qquad  [E]_{\pm }=0, \\
\label{17}
& & [\psi '+{\cal H}\phi 
+({\cal H}'-{\cal H}^2)\frac{\delta q}{q_0'}]_{\pm }=0, 
\qquad [B-E'+\frac{\delta q}{q_0'}]_{\pm }=0.
\end{eqnarray}
This result was generalized for spatially curved backgrounds and 
for the other types of perturbations in Ref.~2. If the surface of transition 
is a surface of constant time, the matching conditions in the synchronous 
gauge are:
\begin{equation}
\label{18}
[h]_{\pm}=[h']_{\pm }=[h_l]_{\pm }=[h_l']_{\pm }=0.
\end{equation}
For a sharp transition (i.e. if $\gamma $ is discontinuous) the two 
sets are not equivalent. Since it is believed that the transition occurs on 
a surface of constant energy which is not a surface of constant time, using 
the second set of joining conditions in this situation leads to an error. 
However, if the transition is smooth, i.e. $[p]_{\pm }=0$, 
then the two sets are equivalent, see Ref.~2.
\par
We use this result to study the regularisation scheme of 
Grishchuk\cite{G}: Instead of matching inflation 
to the radiation epoch directly, Grishchuk introduced a smooth transition in 
between. Physically, this smooth transition represents the reheating of 
the Universe. It starts at $\eta =\eta _1-\epsilon$ 
(end of inflation) and ends at $\eta =\eta _1+\epsilon$ 
(beginning of radiation). Here, $\epsilon$ is small compared to 
$\eta_1$ because we assume that reheating is a fast process. 
In the limit $\epsilon$ goes to zero, we recover the sharp transition 
considered before and $\gamma (\eta )$ becomes a Heaviside function 
jumping from $(2+\beta )/(1+\beta )$ to $2$. For a smooth 
transition, without taking all the details of the reheating process 
into account, we do not know how the scale factor
(and therefore $\gamma$) evolves between $\eta _1-\epsilon$ and 
$\eta _1+\epsilon$. The idea of Ref.~1 was to assume 
that the function $\gamma (\eta )$ is given by:
\begin{equation}
\label{20}
\gamma(\eta) = {4 + 3 \beta\over 2(1 + \beta)} + {\beta\over 2(1 + \beta)} 
\tanh\left( \eta - \eta_1\over s\right) \ ,
\end{equation}
where $s$ is a parameter controlling the sharpness of the transition. This 
equation holds for inflation and reheating, i.e. for $\eta $ between 
$-\infty $ and $\eta _1+\epsilon$. Formula (\ref{20}) leads to a 
reasonable equation of state $p/\rho=-1+2/3\gamma (\eta )$ and therefore 
gives a reasonable approximation to the real (exact) complicated 
function $\gamma (\eta )$ even if details 
of the reheating process cannot be taken into account in such a 
simple approach.
\par
{}From the previous discussion, it is clear that $\gamma (\eta )$ is 
always continuous, even at $\eta =\eta _1+\epsilon$ where 
the explicit joining was performed in Ref.~1 \footnote{ 
In this paper $\eta _1$ was used instead of $\eta _1+\epsilon$ to denote
the end of reheating. It is very important to distinguish these two events.}. 
This means that $[p]_{\pm}$ vanishes. In Ref.~3, Deruelle and 
Mukhanov criticized the calculations 
done in Ref.~1 by means of the smooth transition described before, 
arguing that the junction conditions were not taken into account 
properly. We have shown that for continuous pressure the two sets of 
matching conditions are equivalent. Therefore the claim of Deruelle and 
Mukhanov is not appropriate. For a smooth transition, the matching 
conditions used by Grishchuk are perfectly justified since they 
coincide with the ones derived in Ref.~3. The argument of 
Deruelle and Mukhanov would be relevant 
if the transition was sharp and $\gamma (\eta )$ discontinuous 
at $\eta =\eta _1+\epsilon$. 

Even the simple form (\ref{20}) is too complicated to allow a 
direct integration of the equation of motion for $\mu $. Nevertheless, we 
can follow the 
evolution of $\mu $ through inflation and reheating. For 
$\eta <\eta _1-\epsilon$, $\gamma (\eta )$ is a constant and the solution 
for $\mu $ is given by Eq.~(\ref{10}). The value of $\mu(\eta )$ just 
before reheating can be easily computed:
\begin{equation}
\label{22}
\mu (\eta_1 - \epsilon) \simeq {A_1^{\rm (dp)} \over 2^{\beta+\frac12}
\Gamma(\beta +\frac32) } [k(\eta_1 - \epsilon)]^{\beta + 1} \simeq
{A_1^{\rm (dp)} \over 2^{\beta+\frac12}
\Gamma(\beta +\frac32) } (k\eta_1)^{\beta + 1} \ ,
\end{equation}
because $k\eta_1 \ll 1$ and $\epsilon \ll \eta_1$. Between 
$\eta _1-\epsilon$ and $\eta _1+\epsilon$ the function $\gamma (\eta )$ is 
no longer a constant and the solution (\ref{10}) can no longer be 
used. In order to evolve 
$\mu$ through the reheating transition we use the 
superhorizon solution of Eq.~(\ref{mu}), $\mu \sim a\sqrt{\gamma}$, to obtain 
\begin{equation}
\label{23}
\mu(\eta_1 + \epsilon) \simeq 
\frac{\mu(\eta_1 - \epsilon)}{a(\eta_1 - \epsilon)
\sqrt{\gamma(\eta_1 - \epsilon)}} a(\eta_1 + \epsilon)  
\sqrt{\gamma(\eta_1 + \epsilon)} \simeq \mu(\eta_1 - \epsilon) 
\sqrt{2\over \gamma_{\rm i}} \ ,
\end{equation}
because $a(\eta _1+\epsilon)\approx a(\eta _1-\epsilon)$. This relation 
should be compared to Eq.~(81) and to the relation 
$\mu |_{\eta _1-0}=\mu |_{\eta _1+0}$ below Eq.~(48) of Ref.~1. 
{}From Eq.~(\ref{23}), it is clear that the ratio 
$\mu (\eta _1+\epsilon)/\mu(\eta _1-\epsilon)$ is not $1$ but 
proportional to $1/\sqrt{\gamma_i}$. This factor is 
huge when $\gamma_{\rm i}$ 
is close to $0$ (de Sitter). Therefore the mistake in 
Ref.~1 was not the use of wrong junction conditions but 
the fact that $\mu(\eta )$ was not evolved correctly through 
the reheating transition: $\gamma (\eta _1-
\epsilon)\neq \gamma (\eta _1+\epsilon)$ implies 
$\mu(\eta _1-\epsilon)\neq \mu (\eta _1+\epsilon)$. 
\par
Let us turn to the radiation-matter transition taking place at equality. 
In Ref.~1 the equality transition was treated as a sharp transition. The 
joining conditions (\ref{18}) are not correct in general 
(i.e., for any residual gauge fixing). However, if one specifies
the synchronous gauge in the matter dominated epoch to be the comoving one,
then the joining conditions (\ref{18}) are fine at the equality transition 
\cite{DM}. That is because the density contrast vanishes at the leading 
order [i.e., it is proportional to $(k\eta)^2$]. 
\par
Using the previous results, it is interesting to calculate the ratio 
of the Bardeen potential $\Phi\equiv \phi + [a(B-E')]'/a$ and 
$h_{\rm gw}$ at superhorizon scales today. We obtain (see Ref.~2): 
\begin{equation}
\label{30}
\left. \frac{h_{\rm gw}}{\Phi} \right|_{\rm today}=\frac{10\sqrt{2}}{3}
\frac{\rm A_1^{(gw)}}{\rm A_1^{(dp)}}
\frac{\mu (\eta _1-\epsilon )}{\mu (\eta _1+\epsilon )} = 
\frac{10}{3}\sqrt{\gamma_{\rm i}} \ .
\end{equation}
This equation illustrates the importance of the initial conditions and 
the importance of the evolution of density perturbations during 
reheating. For small values of $\gamma_{\rm i}$ the amplitude of scalar 
metric 
perturbations is larger than the amplitude of gravitational waves, which 
implies that the main contribution to the large scale fluctuations in the 
cosmic microwave background is due to density fluctuations. 

\section*{Acknowledgments}

D.J.S. thanks the Alexander von Humboldt foundation for a fellowship.

\newpage

{\bf Titre en fran\c cais:} Amplitude des perturbations cosmologiques 
plus grandes que l'horizon.

\vspace{2cm}

{\bf R\'esum\'e en fran\c cais:}
\par
Nous \'etudions l'influence du 'reheating' sur les perturbations de densit\'e 
et les ondes gravitationnelles plus grandes que l'horizon. Nous corrigeons 
de fausses affirmations faites \`a propos du raccordement des perturbations 
lors des transitions cosmologiques et \`a propos de leur quantification.

\end{document}